\documentclass{svmult6}
\usepackage{graphicx}

\begin{document}

\chapter{Delaunay Recovery of Cosmic Density and Velocity Probes}

\chapterauthors{W.E.~Schaap \& R.~van de Weygaert\footnote{Kapteyn Institute, PO Box 800,
    9700 AV Groningen, The Netherlands}}
\vspace*{-2.6cm}
\begin{abstract}
Optimally resolved one-dimensional density and velocity profiles through 
cosmological N-body simulations are constructed by means of the 
Voronoi-Delaunay tessellation reconstruction technique. In a fully 
self-adaptive fashion a strikingly detailed view of the density 
features and the corresponding cosmic motions is recovered.
\end{abstract}

\vspace*{.25cm} In essence, N-body simulations of cosmic structure
formation are supposed to represent a discrete sampling of underlying
continuous density and dynamical fields. The recovery of the
corresponding continuous fields is a less than trivial exercise. They
are often distorted by manipulated, user-dependent and therefore
biased reconstruction schemes. This makes it in particular cumbersome
to deal self-consistently with the characteristically multi-scale
hierarchical nature of cosmological density fields. As significant is
the failure to recover crucial structural aspects of the salient
and frequently sharply defined anisotropic -- filamentary and
wall-like -- patterns in the cosmic matter distribution.

Recently, Schaap \& van de Weygaert \cite{sch00} have developed a
fully self-adaptive and unbiased method to reconstruct density and
related dynamic fields from a discrete and in general nonuniformly
sampled set of point locations. It is based on the stochastic
geometric concept of Voronoi/Delaunay tessellations and forms an
elaboration on the formalism first proposed by Bernardeau \& van de
Weygaert \cite{ber96} for the case of assessing the statistical
properties of cosmic velocity fields.

The application of the method to a large $256^3$ GIF N-body simulation
(LCDM, $141.3h^{-1}\hbox{Mpc}$, courtesy: S.~White) \cite{kau99,sch00}
provides a beautiful illustration of its sizeable promise. The top
panel of fig.~1 presents the particle distribution in a slice through
this simulation. The corresponding density field determined through
the Delaunay technique is shown in the adjacent panel. Notice how much
better than the saturated particle plot this density field manages to
elucidate the wealthy and detailed structural features present in this
cosmic volume, superbly rendering its high density contrasts.

While the image of the density field already provides evidence of its
operation, it is through the objective assessment of the density
profile along the central axis of the slice that the success of the
method is fully manifested (solid line, lower panel). Evidently, the
Delaunay technique yields a faithful representation of the density
field over an impressive dynamic range, encompassing gently varying
and extended low-density regions as well as the high density contrasts
found in compact objects, be it either condensed clumps or the
flattened dimension(s) of filaments and walls. Even more compelling is
the correlation with the corresponding velocity field along the same
line (dashed line). Largely superseding the poor velocity resolution
in the conventionally shotnoise-dominated void regions it succeeds in
reproducing the matter depleting {\it super-Hubble} like peculiar
velocity flows (e.g. void at $\approx 123h^{-1}\hbox{Mpc}$). Even more
striking are the sharp velocity transitions encountered at the
location of high density peaks, indicating the large induced infall
motions along various directions towards these features (e.g. the peak
at $\approx 7h^{-1}\hbox{Mpc}$).

\begin{figure}[t]
\vspace*{8cm}
\vspace*{-.4cm}\caption{Slice through a GIF N-body simulation (top), the Delaunay recovered 
density (center) and density and velocity profiles along the central line 
(bottom). (We are grateful to S.~White for initiating these calculations.)}
\end{figure}

\end{document}